\documentclass[a4paper,11pt]{article}
\usepackage{pos}

\usepackage[symbol]{footmisc}
\usepackage[english]{babel}
\usepackage{graphicx}
\usepackage{graphics}
\usepackage{braket}
\usepackage{bbold}
\usepackage{amsmath}
\usepackage{nicefrac}
\usepackage{dcolumn}
\usepackage{bm}
\usepackage{slashed}
\usepackage{datetime}
\usepackage{mciteplus}
\usepackage{multirow}
\usepackage{bbold}
\usepackage{siunitx}
\usepackage{booktabs}
\usepackage{color, soul}
\usepackage[usenames,dvipsnames]{xcolor}
\usepackage{float}
\usepackage[utf8]{inputenc}
\usepackage[normalem]{ulem}
\usepackage{mathtools}
\usepackage{setspace}

\renewcommand{\thefootnote}{\fnsymbol{footnote}}

\newcommand{\LQCD}{\Lambda_{\rm QCD}}

\newcommand{\NLLm}{{\rm NLL/NLO^-}}

\newcommand{\DY}{\Delta Y}

\title{Towards high-energy Higgs+jet distributions at NLL matched to NLO}
\ShortTitle{Towards Higgs+jet distributions at NLL matched to NLO}

\author*[a]{Francesco Giovanni Celiberto}
\author[b,c]{Luigi Delle Rose}
\author[d]{Michael Fucilla}
\author[b,c]{Gabriele Gatto}
\author[b,c]{Alessandro Papa}

\affiliation[a]{Universidad de Alcalá (UAH), Departamento de Física y Matemáticas, Campus Universitario, Alcalá de Henares, E-28805, Madrid, Spain}

\affiliation[b]{Dipartimento di Fisica, Università della Calabria, Arcavacata di Rende, I-87036, Cosenza, Italy}

\affiliation[c]{
INFN, Gruppo Collegato di Cosenza, Arcavacata di Rende, I-87036, Cosenza, Italy}

\affiliation[d]{Université Paris-Saclay, CNRS/IN2P3, IJCLab, 91405, Orsay, France}

\emailAdd{francesco.celiberto@uah.es}
\emailAdd{luigi.dellerose@unical.it}
\emailAdd{michael.fucilla@ijclab.in2p3.fr}
\emailAdd{gabriele.gatto@unical.it}
\emailAdd{alessandro.papa@fis.unical.it}

\abstract{We report progress on the study of the inclusive semi-hard hadroproduction of a Higgs+jet system at LHC and FCC collision energies. We describe a prototype matching procedure aimed at combining NLO fixed-order computations via POWHEG, with the NLL resummation of energy logarithms from JETHAD. We present preliminary analyses on assessing the weight of systematic uncertainties, such as the ones coming from finite top- and bottom-quark masses.}

\FullConference{The European Physical Society Conference on High Energy Physics (EPS-HEP2023)\\
 21-25 August 2023\\
Hamburg, Germany\\}

\begin{document}
\maketitle

\section{Introduction}
\label{sec:introduction}

The discovery of the Higgs boson at the LHC set the stage for a new season of precision analyses of the Standard Model and of searches for evidence of New Physics.
To this extent, a precise description of Higgs production \emph{via} gluon fusion in perturbative Quantum Chromodynamics (QCD) is needed~\cite{Dawson:1990zj,Djouadi:1991tka}.
Higher-order computations are the common basis for accurate studies of Higgs production within the standard \emph{collinear factorization}.
Nonetheless, a proper description of Higgs-sensitive final states in kinematic ranges accessible at the LHC as well as the future FCC rely upon the \emph{all-order resummation} of logarithms which can be large enough to hamper the convergence of the perturbative series.
In this work we focus on the \emph{semi-hard} regime~\cite{Gribov:1983ivg,Celiberto:2017ius,Hentschinski:2022xnd}, where the rigorous scale hierarchy, $\LQCD \ll \{Q_i\} \ll \sqrt{s}$, brings to the rise of large energy logarithms.
Here, $\{Q_i\}$ is a set of process-related hard scales, while $\sqrt{s}$ is the center-of-mass energy.
The Balitsky--Fadin--Kuraev--Lipatov (BFKL) formalism~\cite{Fadin:1975cb,Balitsky:1978ic} is the well-established reference tool to perform the all-order resummation of these logarithms with leading-logarithmic (LL) and next-to-leading logarithmic (NLL) accuracy.
The BFKL resummation and its extension to saturation offer us a chance to unravel the small-$x$ gluon density in the proton~\cite{Bacchetta:2020vty,Arbuzov:2020cqg,Celiberto:2021zww,Celiberto:2022omz,Amoroso:2022eow,Bolognino:2018rhb,Bolognino:2021niq,Celiberto:2019slj,Peredo:2023oym,Taels:2022tza,Caucal:2023nci}.
Golden channels to probe BFKL and high-energy QCD in hadron scattering are the semi-inclusive detections of two objects with large transverse masses and being separated by a large rapidity distance, $\DY$.
A proper description of these two-particle processes requires the development of a \emph{multilateral} approach, where both high-energy and collinear dynamics are at work. Pursuing this goal, a \emph{hybrid high-energy and collinear factorization} (HyF) was constructed~\cite{Celiberto:2020tmb,Bolognino:2021mrc,Colferai:2023hje} (see also~\cite{Deak:2009xt,vanHameren:2022mtk,Bonvini:2018ixe,Bonvini:2018iwt,Silvetti:2022hyc} for single-particle emissions).
HyF partonic cross sections are cast as convolutions between two  process-dependent impact factors (or emission functions) and the universal NLL BFKL Green's function (corresponding to the Sudakov radiator in soft-gluon resummations).
Emission functions are in turn factorized \emph{via} a convolution of collinear parton distribution functions (PDFs) with singly off-shell coefficient functions.
The state-of-the-art accuracy of HyF is NLL/NLO, which means that, for the process under investigation, the required coefficient functions must be computed at fixed next-to-leading order (NLO) accuracy. 
Contrarily, one must go with a partial next-to-leading accuracy, labeled as $\NLLm$, where the Green's function is at NLL, one coefficient function is at NLO, and the other one is at LO.
Excellent probe channels for the HyF formalism are: Mueller--Navelet jet tags~\cite{Ducloue:2013hia,Ducloue:2013bva,Celiberto:2015yba,Celiberto:2015mpa,Celiberto:2016ygs,Caporale:2018qnm,Celiberto:2022gji}, multi-jet detections~\cite{Caporale:2016soq,Caporale:2016zkc,Caporale:2015int,Caporale:2016xku,Celiberto:2016vhn}, Drell--Yan pairs~\cite{Motyka:2016lta,Celiberto:2018muu,Golec-Biernat:2018kem,Taels:2023czt}, light~\cite{Celiberto:2016hae,Celiberto:2017ptm,Celiberto:2016zgb,Bolognino:2018oth,Bolognino:2019yqj,Bolognino:2019cac,Celiberto:2020rxb,Celiberto:2022kxx,Ivanov:2012iv,Fucilla:2023mkl} as well as singly heavy-quark~\cite{Celiberto:2017nyx,Bolognino:2019yls,Bolognino:2019ccd,AlexanderAryshev:2022pkx,Celiberto:2021dzy,Celiberto:2021fdp,Celiberto:2022zdg,Celiberto:2022keu,Anchordoqui:2021ghd,Feng:2022inv} hadrons, quarkonia~\cite{Boussarie:2017oae,Chapon:2020heu,Celiberto:2022dyf,Celiberto:2023fzz}, and exotic bound states~\cite{Celiberto:2023rzw}.
In this study we consider the inclusive Higgs-plus-jet reaction~\cite{Celiberto:2020tmb,Celiberto:2023rtu,DelDuca:1993ga}, which was addressed \emph{via} the next-to-NLO perturbative QCD~\cite{Chen:2014gva,Boughezal:2015dra} and by the next-to-NLL transverse-momentum resummation~\cite{Monni:2019yyr}.
We present the POWHEG+JETHAD method, a new \emph{matching} procedure aimed at combining NLO fixed-order results with the resummation of NLL energy logarithms.

\section{Towards Higgs-plus-jet production at NLL/NLO}
\label{sec:matching}

\begin{figure*}[!t]
\centering

\includegraphics[scale=0.36,clip]{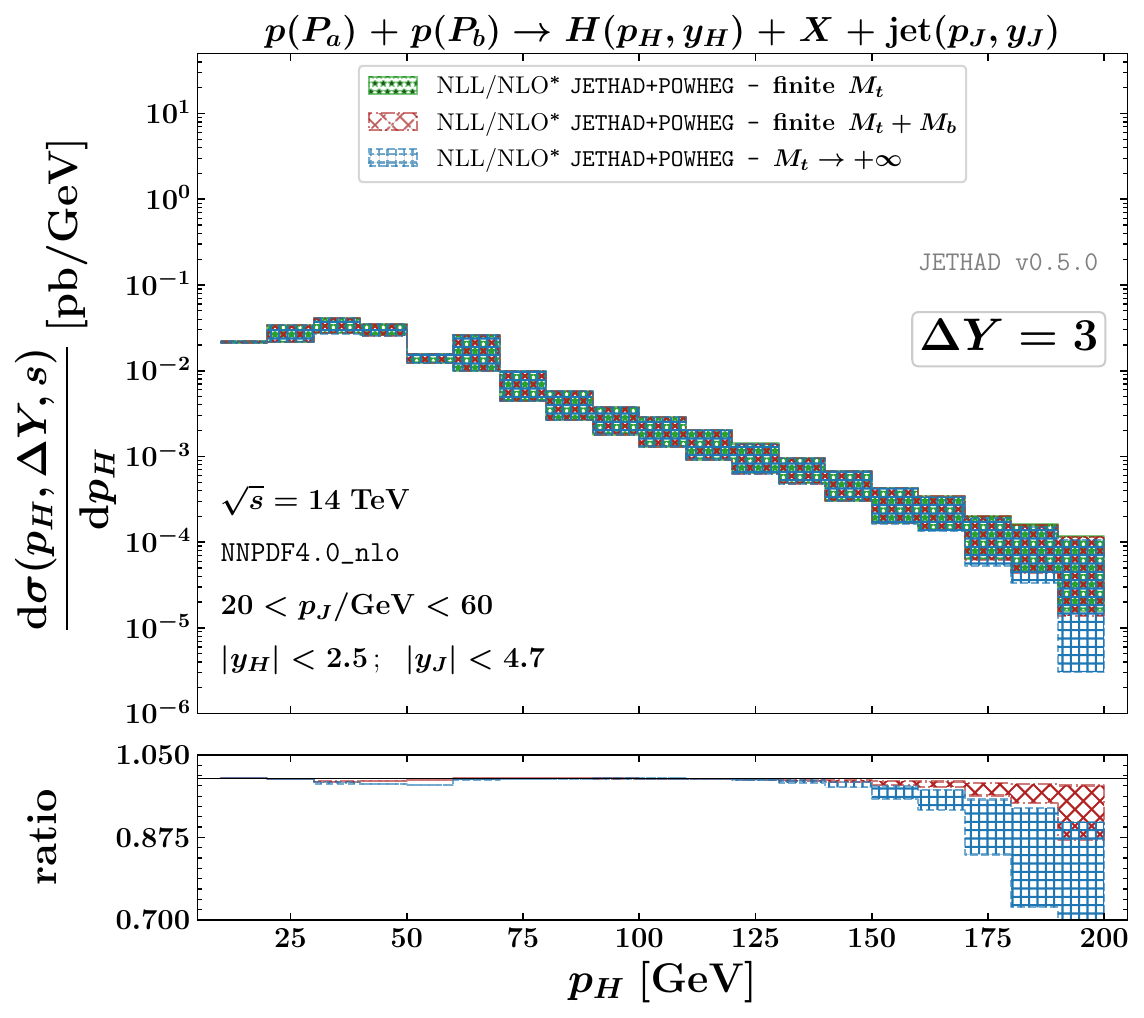}
 \hspace{0.30cm}
\includegraphics[scale=0.36,clip]{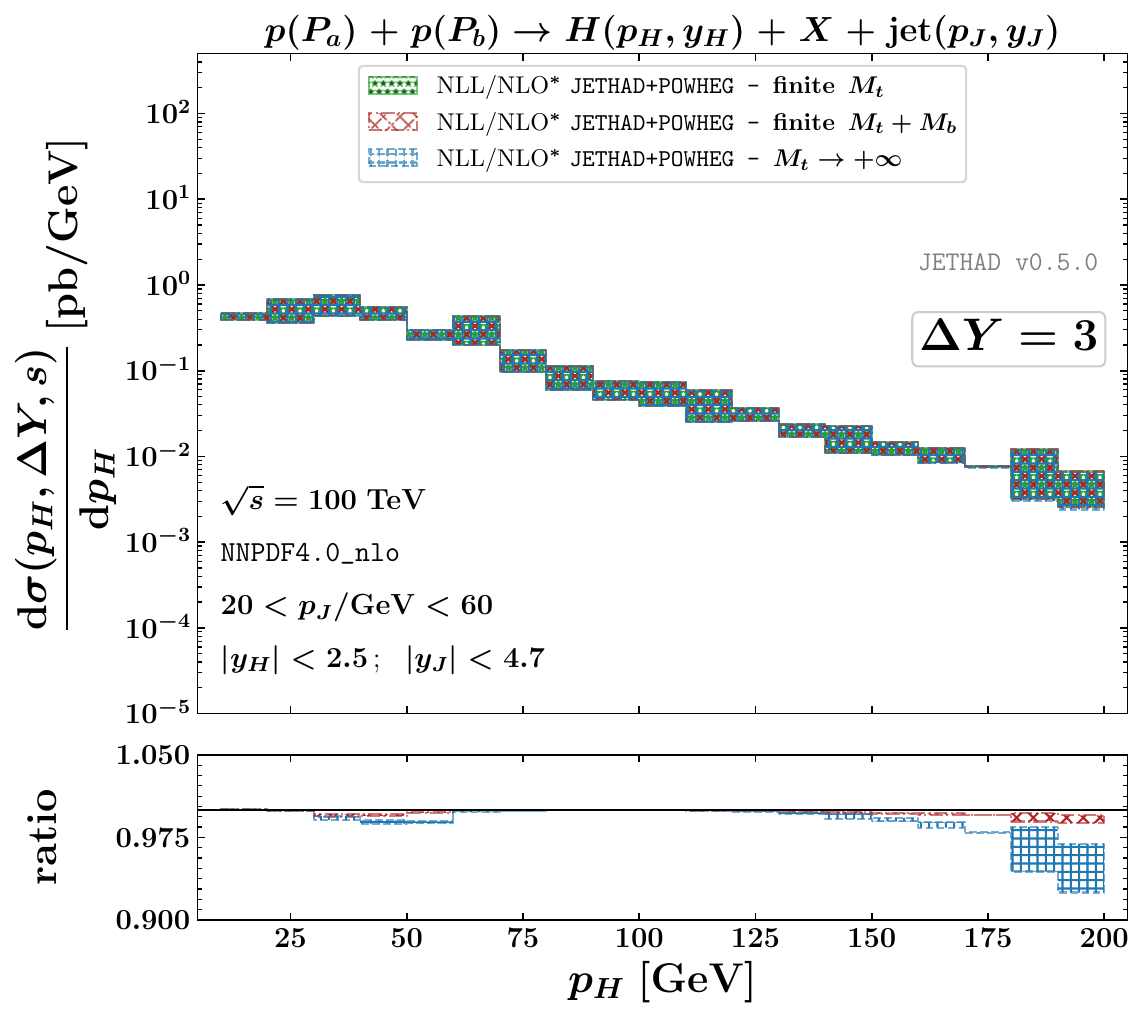}

\caption{Higgs-plus-jet transverse-momentum rates at $14$ (LHC, left) and $100$ (FCC, right) collision energies.
Shaded bands exhibit $\mu_{R,F}$ variation in the $1 < C_{\mu} < 2$ range. Text boxes are for kinematic cuts.
}

\label{fig:pT_mcrs}
\end{figure*}

Recent HyF analyses on the Higgs transverse-momentum ($p_H$) spectrum in inclusive Higgs-plus-jet detections at the LHC~\cite{Celiberto:2020tmb} and the future FCC~\cite{Celiberto:2023rtu} have shown a fair stability under higher-order corrections and energy-scale variation.
At the same time, strong deviations of HyF results from pure fixed-order predictions emerged, their size being up to two orders of magnitude larger when $p_H \gtrsim 120$~GeV.
To cure this problem, we have developed a pioneering \emph{matching} procedure between the NLO fixed-order signal and the NLL resummation, which implements an exact removal, within the $\NLLm$ accuracy, of the corresponding \emph{double counting}.
Since the NLO forward-Higgs impact factor~\cite{Celiberto:2022fgx,Hentschinski:2020tbi} still has to be embodied in the JETHAD technology~\cite{Celiberto:2020wpk,Celiberto:2022rfj,Celiberto:2023fzz}, we will go with a $\NLLm$ description.
Our matching scheme is the following~\cite{Celiberto:2023uuk_article,Celiberto:2023eba_article}

\begin{equation}
\label{eq:matching}
\begin{split}
 \hspace{-0.155cm}
 \underbrace{{\rm d}\sigma^{{{\rm NLL/NLO}}^{\boldsymbol{-}}}(\Delta Y, \varphi, s)}_{\text{\colorbox{OliveGreen}{\textbf{\textcolor{white}{NLL/NLO$^{\boldsymbol{-}}$}}} {\tt POWHEG+JETHAD}}} 
 = 
 \underbrace{{\rm d}\sigma^{\rm NLO}(\Delta Y, \varphi, s)}_{\text{\colorbox{gray}{\textcolor{white}{\textbf{NLO}}} {\tt POWHEG} w/o PS}}
 +\; 
 \underbrace{\underbrace{{\rm d}\sigma^{{{\rm NLL}}^{\boldsymbol{-}}}(\Delta Y, \varphi, s)}_{\text{\colorbox{red}{\textbf{\textcolor{white}{NLL$^{\boldsymbol{-}}$ resum}}} (HyF)}}
 \;-\; 
 \underbrace{\Delta{\rm d}\sigma^{{{\rm NLL/NLO}}^{\boldsymbol{-}}}(\Delta Y, \varphi, s)}_{\text{\colorbox{orange}{\textbf{NLL$^{\boldsymbol{-}}$ expanded}} at NLO}}}_{\text{\colorbox{NavyBlue}{\textbf{\textcolor{white}{NLL$^{\boldsymbol{-}}$}}} {\tt JETHAD} w/o NLO$^{\boldsymbol{-}}$ double counting}}
 \,.
\end{split}
\end{equation}

A given observable, ${\rm d}\sigma^{{{\rm NLL/NLO}}^{\boldsymbol{-}}}$, matched at $\NLLm$ (green) by the POWHEG+JETHAD method, reads as a sum of the NLO fixed-order contribution (gray) from POWHEG~\cite{Nason:2004rx,Campbell:2012am,Hamilton:2012rf,Banfi:2023mhz,Bagnaschi:2023rbx} (without \emph{parton shower} (PS)~\cite{Alioli:2022dkj,Alioli:2023har,Buckley:2021gfw,vanBeekveld:2022zhl,vanBeekveld:2022ukn,FerrarioRavasio:2023kyg,Black:2022cth}) with the $\rm NLL^-$ resummed term (blue) from JETHAD. The latter is the $\rm NLL^-$ HyF resummed part (red) minus the $\rm NLL^-$ expanded (orange) at NLO, namely without double counting.
As an extension of our preliminary study~\cite{Celiberto:2023uuk_article,Celiberto:2023eba_article}, we present in Fig.~\ref{fig:pT_mcrs} results on the $p_H$ distribution 14~TeV~LHC (left) and 100~TeV~FCC (right), and for $\DY = 3$.
Our analysis shows that progressively including effects of finite top (green) and bottom (firebrick) mass corrections has an almost negligible impact on the distribution when $p_H \lesssim 120$~GeV (except for the peak region). Then, it increases with $p_H$ up to reach almost $70\%$, thus indicating that mass corrections become more and more important in the large $p_H$ tail, as expected.

\section{Paving the way towards precision}
\label{sec:conclusions}

With the aim of combining NLO fixed-order results with the NLL high-energy resummation, we presented a first and prototype version of a matching procedure, based on the POWHEG~\cite{Nason:2004rx,Campbell:2012am,Hamilton:2012rf,Banfi:2023mhz,Bagnaschi:2023rbx} and JETHAD~\cite{Celiberto:2020wpk,Celiberto:2022rfj,Celiberto:2023fzz} technologies.
Future studies are needed to: $1)$ include NLO contributions from the Higgs coefficient function~\cite{Celiberto:2022fgx,Hentschinski:2020tbi}, $2)$ complete our analysis on quark-mass contributions~\cite{Jones:2018hbb,Bonciani:2022jmb}, $3)$ compare our results with PS~\cite{Alioli:2022dkj,Alioli:2023har,Buckley:2021gfw,vanBeekveld:2022zhl,vanBeekveld:2022ukn,FerrarioRavasio:2023kyg} and HEJ~\cite{Andersen:2022zte,Andersen:2023kuj} related ones.

\section*{Acknowledgments}
\label{sec:acknowledgments}

This work was supported by the Atracci\'on de Talento Grant n. 2022-T1/TIC-24176 of the Comunidad Aut\'onoma de Madrid, Spain, and by the INFN/QFT@COLLIDERS Project, Italy. M.~F. is supported by Agence Nationale de la Recherche under the contract ANR-17-CE31-0019.

\begingroup
\setstretch{0.6}
\bibliographystyle{bibstyle}
\bibliography{biblography}
\endgroup

\end{document}